# Investigation of Spin-Pumping and -Transport in the Ni$_{80}$Fe$_{20}$/Pt/Co Asymmetric Trilayer

Shilpa Samdani[1], Yaqi Rong[2], Birte Coester[3], Amit Kumar Shukla[1], Lew Wen Siang[3], Yumeng Yang[2], and Rajdeep Singh Rawat[1*]

[1] *Natural Sciences and Science Education, National Institute of Education, Nanyang Technological University, Singapore 637616*
[2] *Shanghai Engineering Research Center of Energy Efficient and Custom AI IC, School of Information Science and Technology, ShanghaiTech University, Shanghai, 201210, China*
[3] *School of Physical and Mathematical Sciences, Nanyang Technological University, Singapore 637371*

Ferromagnet$_1$/Non-magnetic Metal/Ferromagnet$_2$ (FM$_1$/NM/FM$_2$) trilayers have garnered considerable attention because of their potential in spintronic applications. A thorough investigation of the spin transport properties of these trilayers is therefore important. Asymmetric trilayers, particularly those including Platinum (Pt) as a spacer are less explored. Pt mediates exchange coupling between the two FM layers and thus offers a unique platform to investigate the spin-transport properties under indirect exchange coupling conditions through the spin-pumping mechanism. We study the static and dynamic magnetic properties of Ni$_{80}$Fe$_{20}$/Pt(t)/Co trilayer system through vibrating sample magnetometry (VSM) and spin-pumping based on ferromagnetic resonance (FMR) spectroscopy by varying the Pt spacer thickness. Though a powerful method for characterizing the dynamic magnetic properties of FM layers, FMR is seldom the only technique used for investigating spin transport characteristics of asymmetric trilayers. Our analytical focus on the acoustic mode, facilitated by the distinct magnetizations of the Ni$_{80}$Fe$_{20}$ and Co layers, allows for the isolation of individual layer resonances. The derived spin-pumping induced damping ($\alpha_{sp}$) of the Ni$_{80}$Fe$_{20}$ and Co layers reveals a direct dependence on the Pt spacer thickness. Furthermore, fitting of the weighted average of the damping parameters to the $\alpha_{sp}$ of acoustic mode reveals that the observed FMR spectra is indeed a result of the in-phase precession of the magnetizations in two FM layers. The extracted effective spin-mixing conductance ($g_{eff}^{\uparrow\downarrow}$) varies with the FM/NM interface, specifically $1.72 \times 10^{19}\ m^{-2}$ at the Ni$_{80}$Fe$_{20}$/Pt and $4.07 \times 10^{19}\ m^{-2}$ at the Co/Pt interface, indicating a strong correlation with interfacial characteristics. Additionally, we deduce the spin diffusion length in Pt to be between 1.02 and 1.55 nm and calculate the interfacial spin transparency ($T_{in}$) and spin current densities, highlighting significant disparities between the Ni$_{80}$Fe$_{20}$/Pt and Co/Pt interfaces. This detailed analysis enhances our understanding of spin transport in Ni$_{80}$Fe$_{20}$/Pt/Co trilayers. It offers insights important for advancing spintronic device design and lays the groundwork for future theoretical investigations of trilayer system.

## I. INTRODUCTION

Trilayer systems consisting of two ferromagnetic (FM) layers separated by a non-magnetic (NM) layer have attracted the interest of magnetism and spintronics research community since the discovery of the Giant Magneto-Resistance (GMR) effect in the 1990s. Traditionally, trilayers have been studied for applications in magnetic recording and random memories. In recent years, a renewed interest in these trilayers is seen in emerging areas of spintronics studies such as field-free spin-orbit torque switching[1], spin-torque microwave nanodevices[2-4] and synthetic antiferromagnetic devices[5, 6]. Understanding the transfer of pure spin current[7, 8] in the trilayers, in other words the 'spin transport', is important for the development of low-power and high-speed spintronic devices. Investigating the magnetization dynamics of the FM layer makes it possible to gain direct insight into important spin transport properties such as the spin-diffusion length ($\lambda_{sd}$) and the spin-mixing conductance ($g_{\uparrow\downarrow}$). $\lambda_{sd}$ is the characteristic length which the pure spin current traverses in the NM layer before dissipating via spin-flip processes while $g_{\uparrow\downarrow}$ parametrizes the efficiency of spin transport across FM/NM interface[9].

In the case of a symmetric coupled trilayer, i.e. when both the FM layers are of the same material (FM/NM/FM), it is straightforward to obtain the value of spin-mixing conductance ($g_{\uparrow\downarrow}$) at the FM/NM interface by studying the collective magnetization dynamics of the coupled FMs, such as Fe/Ag/Fe[10], NiFe/Pt/NiFe[11], etc. However, complexity arises when the trilayer system is asymmetric, i.e. when the two FM layers are different (FM$_1$/NM/FM$_2$), because the two different FM/NM interfaces result in two different values of $g_{\uparrow\downarrow}$. The differing values of $g_{\uparrow\downarrow}$ in turn results in directional asymmetry in the spin current pumped from FM1 and FM2 and transported into the NM spacer layer. Therefore, to correctly describe the spin transport in asymmetric trilayers it is important to quantify the $g_{\uparrow\downarrow}$ at each FM/NM interface.

To date, various approaches including non-local magnetotransport[12], inverse spin-Hall effect[13] and spin-torque ferromagnetic resonance (ST-FMR)[14] have been used to investigate spin transport properties. Lithography-free approaches such as broadband ferromagnetic resonance (FMR) and x-ray detected FMR (X-FMR) based spin-pumping and time resolved magneto-optical Kerr effect (TR-MOKE) offer an easier route to investigate the pure spin current transport. Hitherto, reports on the investigation of spin transport in asymmetric trilayers by such approaches are rather scarce, they include systems such as synthetic antiferromagnetic (SAF) heterostructures, for example, FeCoV/Ru/NiFe[15] and [Co/Pd/Co]$_{FM1}$/[Ru/Ta]$_{NM}$/CoFeB$_{FM2}$[16], spin transfer torque – magnetic tunnel junctions (STT-MTJ)

CoFeB/Ta/NiFe[17], spin-valve Co/Cu/CoFeB[18], CoFe/Cr/NiFe[19] and novel heterostructures CoFe/Bi$_2$Se$_3$/NiFe[20]. We note that, though FMR is a powerful technique that allows extraction of dynamic magnetic properties such as the effective demagnetizing field, interfacial magnetic anisotropy and the damping parameter, it is rare to find studies where broadband FMR is exclusively utilized to probe spin-transport in trilayer configurations. Typically, it is combined with X-FMR for a detailed examination of each FM layer's resonant properties independently. However, the distinct bulk magnetizations of the two different FM layers in an asymmetrical trilayer system provides an opportunity for resonance deconvolution through lineshape analysis. Furthermore, there are very few reports on the spin-pumping in platinum (Pt) in asymmetric trilayer systems, making FM$_1$/Pt/FM$_2$ heterostructure a novel and compelling system for investigation.

At the FM/Pt interface a finite magnetic moment emerges due to direct exchange coupling, affecting up to few atomic layers of Pt[21]. Additionally, in FM/Pt/FM trilayers there exists a static indirect exchange coupling (IEC) between the FMs. The strength of this coupling diminishes exponentially with the increase in the thickness of the Pt layer[22-25]. The presence of IEC fosters the hybridization of the magnetization precession in the two FM layers resulting in two distinct modes: an in-phase (or acoustic) mode and an out-of-phase (or optical) mode. It is observed that damping of the optical mode is higher than the acoustic mode[11, 26-28] rendering it challenging to observe the optical mode in experiments[29, 30]. In coupled FM$_1$/NM/FM$_2$ systems, spin-pumping induces dynamic coupling between the FM layers. This happens as a result of the spin currents being simultaneously ejected from both the FMs and being non-locally dampened[31]. The absorption of the incoming spin current balances the loss of angular momentum when the magnetizations precess in-phase, but this loss is amplified when the precession is out-of-phase. Consequently, the spin-pumping induced damping is dependent on the precessing magnetizations' phase and amplitude[32].

We present a comprehensive study of the static and dynamic magnetic properties alongside the spin transport characteristics within an asymmetric trilayer Ni$_{80}$Fe$_{20}$/Pt(t)/Co system employing broadband FMR spectroscopy and VSM. By varying the thickness of the Pt spacer layer, we aim to modulate and examine the IEC between the FM layers and its subsequent effect on the magnetization dynamics of the two FM layers. The focal point of our investigation is the acoustic mode of the trilayer system, which due to the differing bulk magnetizations of the two FM layers, enables the deconvolution and individual resonance analysis of each FM layer through lineshape analysis. Our methodological approach is similar to that of Omelchenko *et. al.*[11], the distinguishing factor being that we study an asymmetric trilayer system and subsequently extend the analysis to isolate the separate spin-mixing conductances of the two interfaces. Through FMR and VSM, we extracted key parameters such as the damping co-efficient and coupling strength respectively. Our analysis reveals the dependency of spin-pumping induced damping of the specific FM layers and $g_{\uparrow\downarrow}$ on the respective FM/Pt interface. Additionally, we quantified the interfacial spin transparency and spin current density at each FM/Pt interface. The interfacial spin transparency of the trilayer is found to match FM/Pt bilayer and the spin current density is observed to be dependent on the Pt spacer thickness.

## II. EXPERIMENTAL DETAILS

Thin films were deposited at ambient temperature on thermally oxidized single-crystalline Si substrates utilizing magnetron sputtering. The deposition was carried out in Ar atmosphere at chamber base pressure better than $4 \times 10^{-7}$ Torr. The deposited multilayer sequence was as follows: Si substrate//Ta(2)/Ni$_{80}$Fe$_{20}$(6 nm)/Pt(t$_{NM}$)/Co(6 nm)/Ta(2 nm), where the thickness of the Pt layer (t$_{NM}$) was varied across samples – specifically, 0 nm, 0.5 nm, 1.5 nm, 3.5 nm and 7 nm. To simplify discussions, samples are henceforth designated based on their Pt layer thickness as Pt(0), Pt(1.5), Pt(3.5) and Pt(7) respectively. Underlayer and capping layer of Ta were deposited to promote uniform film growth of Ni$_{80}$Fe$_{20}$ and prevent the oxidation of Co layer respectively. Additionally, a control sample using a spacer layer of Cu was fabricated and analysed to underscore the distinctive effects attributable to Pt.

Broadband, field-swept, in-plane magnetic field, ferromagnetic resonance (FMR) measurements were conducted using Quantum Design VersaLab system with a coplanar waveguide and NanoOsc Phase FMR lock-in detection. Experiments spanned a frequency range of 5 to 20 GHz, with an external DC magnetic field oriented parallel to the sample plane. Samples were saturated at 15 kOe, well above the magnetic field sweep used, to ensure full saturation during measurements. Static magnetic properties were characterized using Vibrating Sample Magnetometry (VSM) on the Quantum Design VersaLab system. These measurements were performed at room temperature, applying fields up to 6 kOe in an in-plane configuration. For samples comprising multiple ferromagnetic (FM) layers, such as the FM$_1$/NM/FM$_2$ structures discussed in this study, the observed hysteresis loops are derived from the collective magnetic responses of both the involved FM layers.

## III. STATIC MAGNETIC PROPERTIES

Figure 1(a) displays the in-plane hysteresis loops that demonstrate the magnetization reorientation transition for a series of Ni$_{80}$Fe$_{20}$(6)/Pt(t)/Co(6) trilayer samples. The thickness, t$_{NM}$, of the Pt spacer varies from 0 to 7 nm. The Pt(0) sample exhibits a square loop with a unique switching field, suggesting that the two ferromagnetic layers are acting synergistically as a single composite entity. This observation is indicative of strong coupling between the FM layers, primarily through direct exchange interactions. The phenomena of IEC in samples with ultrathin NM spacers, such as for Pt(0.5) and Pt(1.5) samples, can be attributed to one or a combination of phenomena, including the Ruddermann-Kittel-Kasuya-Yosida (RKKY) interaction[33], direct exchange coupling via pinholes in the spacer layer[34] and the magnetic proximity effect in Pt[35]. An early magnetization reversal in Pt(3.5), Pt(7) and Cu(7) samples also hints at the presence of magnetostatic (dipole-dipole) coupling[36], prompting a premature reorientation of magnetic moments in the softer FM layer in response to the external magnetic field's directional shifts. The Supplementary Information Section S-I provides a detailed analysis of how the thickness of the Pt spacer layer affects the intrinsic coercive field ($H_{ic}$) and the saturation magnetization ($M_s$).

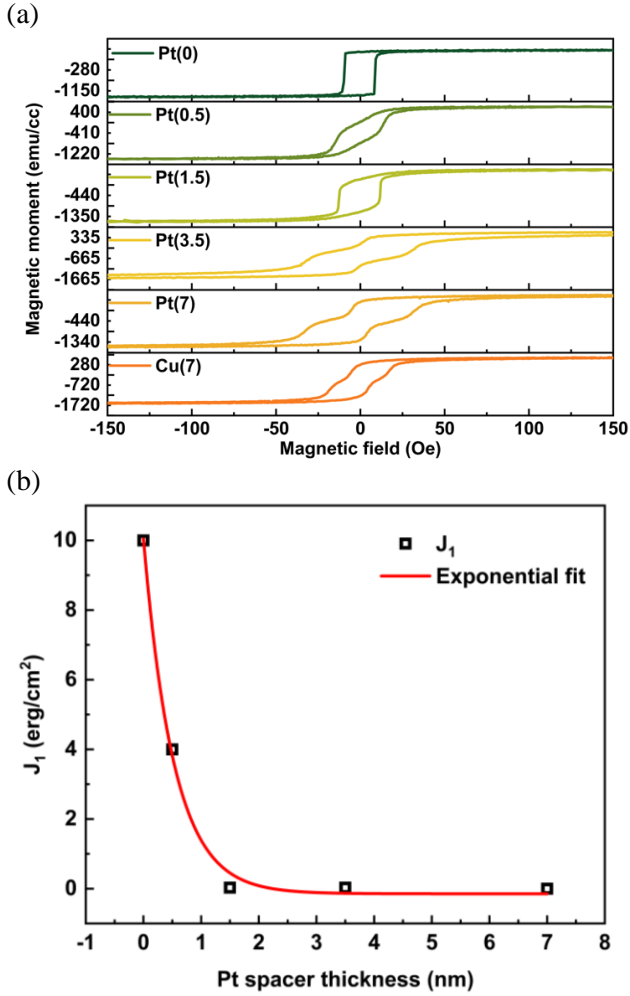

Figure 1: (a) In-plane hysteresis loops of $Ni_{80}Fe_{20}$(6 nm)/NM (t nm)/Co (6 nm) trilayers with NM spacers - Pt(0, 0.5, 1.5, 3 and 7 nm) and Cu(7 nm), and (b) the variation in coupling strength, $J_1$, as a function of the Pt spacer thickness.

The coupling strength ($J_1$) obtained from the macro-spin simulations, refer Supplementary Information Section S-II, is shown in Figure 1(b). It can be modelled as exponentially changing with Pt thickness, with $J_1$ diminishing to nearly zero for $t_{Pt} \geq 1.5\ nm$. This relationship between coupling strength and spacer thickness is consistent with findings in other coupled trilayer systems such as Co/Pt/Co[24], FeNi/Cu/FeCo[37], Fe/Pd/Fe[38], suggesting a universal trend across various material compositions. Specifics regarding the energy equation and simulated M-H loops values are provided in the Supplementary Information Section S-II.

## IV. DYNAMIC MAGNETIC PROPERTIES

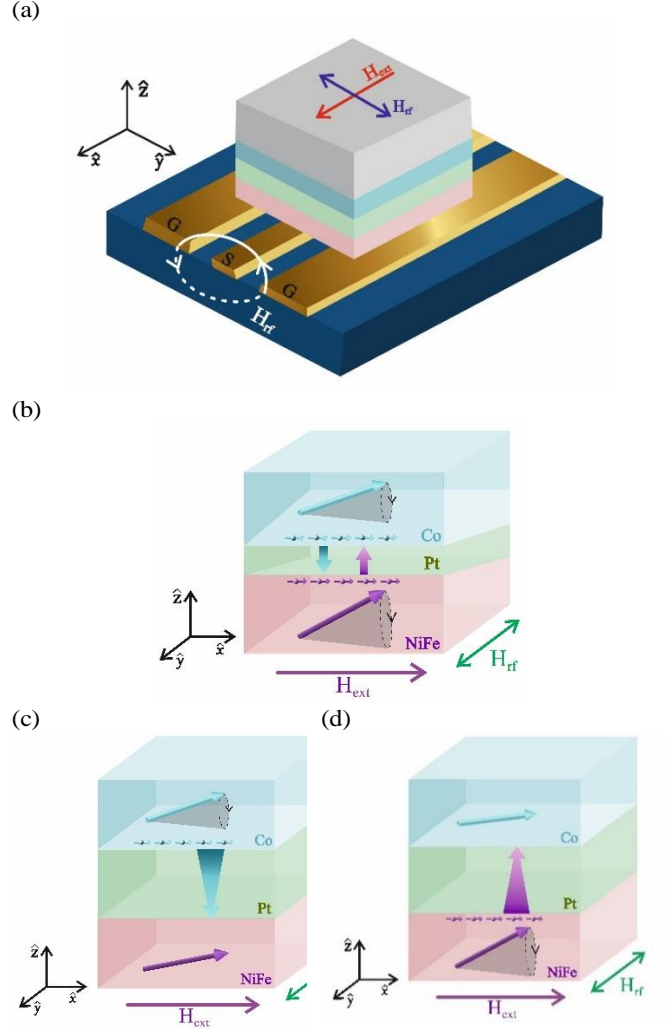

Figure 2: Schematic of (a) configuration of flip-chip co-planar waveguide ferromagnetic resonance spectroscopy with an in-plane magnetic field (red arrow), (b) $FM_1/NM/FM_2$ trilayer of $Ni_{80}Fe_{20}$/Pt/Co showing the in-phase precession of magnetizations of Co and $Ni_{80}Fe_{20}$ layers accompanied by spin-pumping in Pt spacer layer, $t_{Pt} = 0.5$ and 1.5 nm. When, $t_{Pt} = 3.5$ and 7 nm (c) represents spin pumping of Co and (d) represents spin-pumping of $Ni_{80}Fe_{20}$. Cyan and purple arrows in Co and $Ni_{80}Fe_{20}$ layers represent the precessing bulk magnetization. The electron spins in Co (cyan) and $Ni_{80}Fe_{20}$ (purple) represent the ejected spin current. Arrows in the Pt spacer represent the spin current transport in Pt spacer. The spin current pumped by Co and $Ni_{80}Fe_{20}$ is represented by the cyan and purple arrows respectively. Depending on Pt layer thickness the spin current undergoes either ballistic ($t_{Pt} < \lambda_{sd}$) or diffusive ($t_{Pt} > \lambda_{sd}$) transport.

Broadband FMR spectroscopy, is a widely used technique for investigating spin current, enabling the indirect quantification of spin current by analyzing the enhancement in the Gilbert damping parameter[39, 40]. This technique reveals how the IEC influences the effective magnetic field of the coupled FM layers in a trilayer system, manifesting as a shift in the resonance peak position as seen in Figure 3(a).

To elucidate the effect of interlayer coupling on the spin-pumping in our trilayer system, we investigated the dynamic properties of the magnetization by measuring the derivative ($dI/dH$) of the FMR spectra. Figure 2(a) shows a schematic of flip-chip FMR measurement configuration. The in-phase precession of the magnetizations, of Co and $Ni_{80}Fe_{20}$, when FMs are coupled and the independent precession in case of decoupled FMs, alongside the simultaneous spin-pumping into Pt are represented schematically in Figure 2(b-d). The acquired signal is fitted to the derivative of Lorentzian function.

The fitting equation applied in our study is structured as follows.

$$\frac{dI}{dH} = S_1\left[\frac{\frac{\Delta H_1}{2}(H-H_{res,1})}{\left[(H-H_{res1})^2+\left(\frac{\Delta H_1}{2}\right)^2\right]^2}\right] \quad (1)$$

$$+ A_1\left[\frac{\left(\frac{\Delta H_1}{2}\right)^2-(H-H_{res,1})}{\left[(H-H_{res1})^2+\left(\frac{\Delta H_1}{2}\right)^2\right]^2}\right]$$

$$+ offset$$

$$+ S_2\left[\frac{\frac{\Delta H_2}{2}(H-H_{res,2})}{\left[(H-H_{res,2})^2+\left(\frac{\Delta H_2}{2}\right)^2\right]^2}\right]$$

$$+ A_2\left[\frac{\left(\frac{\Delta H_2}{2}\right)^2-(H-H_{res,2})}{\left[(H-H_{res,2})^2+\left(\frac{\Delta H_2}{2}\right)^2\right]^2}\right] + Hm$$

where, $\Delta H_1$ and $\Delta H_2$ represent the resonance linewidths and $H_{res1}$ and $H_{res2}$ represent the resonance magnetic fields of $FM_1$ and $FM_2$ layers, respectively, $m$ is the slope and $S$ and $A$ are amplitudes of the Lorentzian fitting function. Figure 3(a) shows the typical field-swept FMR spectra for the various samples, Ni80Fe20(6)/Pt(t)/Co(6) and Ni80Fe20(6)/Cu(7)/Co(6), all measured at 20 GHz. The experimental data is well-fitted to equation (1) facilitating the precise determination of the resonance field ($H_{res}$) and linewidth ($\Delta H$). When the FM layers are strongly coupled ($t_{Pt} = 0, 0.5, 1.5$ nm) a single FMR peak is observed. This peak exhibits a decrease in intensity and an increase in linewidth, alongside a shift to higher magnetic fields as the spacer layer thickness increases. Conversely, for weakly coupled or decoupled FMs, exemplified by $t_{Pt} = 3.5$ nm, 7 nm and $t_{Cu} = 7$ nm, two distinct resonance peaks emerge, each representing the independent precession of magnetization within the two FM layers. These findings corroborate the insights related to coupling of the two FM layers, gained from examining the static magnetization characteristics in the preceding analysis.

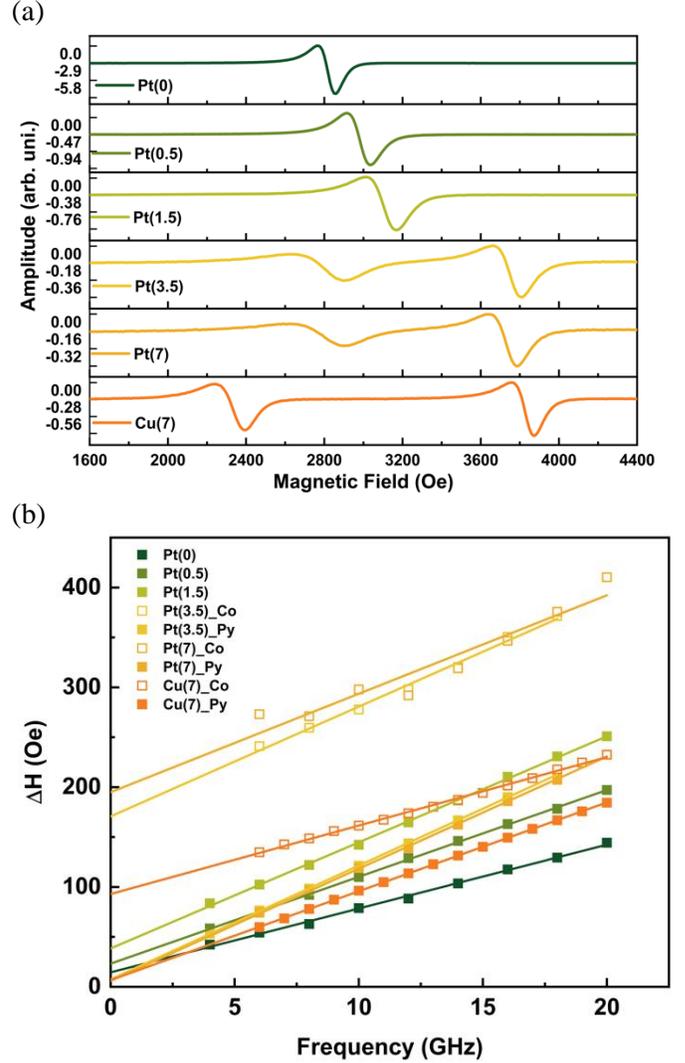

Figure 3: (a) FMR spectra measured at 20 GHz for all the samples, and (b) ferromagnetic resonance linewidth, $\Delta H$, as a function of frequency for varying thickness of Pt spacer, hollow and solid squares represent the linewidth of Co and $Ni_{80}Fe_{20}$ layers respectively. Solid lines represent fit using equation (2) to extract the effective Gilbert damping parameter.

Figure 3(b) shows the dependence of the linewidths ($\Delta H$) of the Co and $Ni_{80}Fe_{20}$ resonance peaks on the FMR excitation frequency across different Pt spacer thicknesses, hollow and solid squares represent the linewidth of Co and $Ni_{80}Fe_{20}$ layers respectively. The linear variation in linewidth as a function of frequency for all the samples indicates the intrinsic origin of damping. Conversely, a non-linear trend is indicative of enhancement in damping due to two-magnon scattering which is not observed in our case. As such, we dismiss extrinsic linewidth contributions arising from two-magnon scattering. From the linear correlation between the $\Delta H$ and frequency, we deduce the effective Gilbert damping parameter ($\alpha_{eff}$) and the linewidth broadening due to inhomogeneities in the film ($\Delta H_0$)[41, 42]:

$$\Delta H = \frac{4\pi\alpha_{eff}}{\gamma}f + \Delta H_0 \quad (2)$$

$\Delta H_0$ also referred to as the zero-frequency broadening is the lowest for Ni$_{80}$Fe$_{20}$ in Pt(7) sample at ~6 Oe. It increases with $t_{Pt}$ while reverse trend is seen in $\Delta H_0$ Co peak as in trilayer with Pt(7) ~194 Oe and 30 Oe for Pt(0) sample. These values are on the higher side as compared to other report of Ni$_{80}$Fe$_{20}$ symmetric trilayer system with Pt spacer[11]. Solid lines in Figure 3(b) represent linear fits to the data. Using equation (2) we determine the effective Gilbert damping parameter, which subsequently allows us to extract the spin-pumping induced damping as shown later in Figure 4.

Further analysis of the FMR spectra concerning the effective magnetization and interfacial anisotropy is discussed in Supplementary Information Section S-III.

We have thus investigated the features and origin of the obtained FMR spectra. We now discuss the main results related to the dynamic magnetic properties of each FM layer and spin transport across each FM/NM interface, starting with the relaxation dynamics at each FM/Pt interface.

## V. SPIN-PUMPING AND SPIN TRANSPORT

Spin-pumping refers to the phenomenon of the transfer of spin current by a precessing magnetization to the neighbouring NM layer because of the non-equilibrium spin accumulation in the FM layer. This process is an outcome of the dynamic and coherent precession of the magnetization around the effective magnetic field (H$_{eff}$) in a FM layer. H$_{eff}$ includes the external magnetic field, crystal anisotropy and demagnetizing fields. The enhancement in the damping of the amplitude of precession of magnetization in the FM layer is understood to be a direct consequence of this spin-pumping phenomenon[39]. The damping of the precessing magnetization is enhanced when the spin current that leaks out of the FM layer dissipates via spin-flip processes in the NM layer after travelling a distance greater than the spin-diffusion length ($\lambda_{sd}$) of the NM layer. However, when $t_{NM} < \lambda_{sd}$, spin accumulation at the FM/NM interface leads to a backflow of the spin current in the FM. Thus, the total spin current at FM/NM interface can be described as a sum of spin-current due to spin-pumping from FM ($I_s^{pump}$) into the NM and backflow spin current from NM into the FM ($I_s^{back}$). In coupled trilayers systems, FM$_1$/NM/FM$_2$, the backflow spin current at the FM$_1$/NM interface also includes spin currents pumped out of FM$_2$ and vice versa. Furthermore, the damping of the precessing magnetization in FM$_1$ (FM$_2$) also depends on the phase of the spin current pumped out of FM$_2$ (FM$_1$). We reserve further discussion on the phase difference of precessing magnetization and its effect on damping till Section VI.

The intrinsic spin-mixing conductance ($g_{int}^{\uparrow\downarrow}$) of an FM/NM interface is a measure of the inherent efficiency of the FM/NM interface in facilitating the transfer of spin angular momentum from the FM to the NM layer, independent of the spin backflow. $g_{int}^{\uparrow\downarrow}$ is determined from the variation of spin-pumping induced damping ($\alpha_{sp}$) as a function of NM thickness ($t_{NM}$). Spin-mixing conductance correlates spin-pumping to the intrinsic properties of the FM and the experimentally measurable effective Gilbert damping parameter, which as outlined earlier, is directly affected by the phenomenon of spin-pumping. It is given as follows[43]:

$$\alpha_{sp} = g_{int}^{\uparrow\downarrow}\frac{g\mu_B\left(1-e^{(-t_{NM})/\lambda_{sd}}\right)}{4\pi M_s t_{FM}} \quad (3)$$

where, all symbols hold their usual meaning. In FM/NM systems the exponential term in the above equation describes the spin diffusion in NM and accounts for the attenuation of spin current as it diffuses through the NM material. To realistically model the spin transport phenomena, the exponential decay term for $t_{NM} < \lambda_{sd}$, is typically multiplied by a factor of 2 that signifies the distance traversed by the spins after reflection from the NM/air interface, which is assumed to be a perfect spin-reflector[44]. However, considering the earlier provided physical explanation of the spin dynamics at the interface in a trilayer, we exclude the factor of 2 from our analysis of FM$_1$/NM/FM$_2$ system, operating under the assumption that the FM layers act as a perfect spin-sink.

On the other hand, when accounting for the spin backflow, the efficiency of spin-pumping across the FM/NM interface is given by the effective complex spin-mixing conductance ($g_{eff}^{\uparrow\downarrow}$) per unit area of the interface. It is expressed as, $g_{eff}^{\uparrow\downarrow} = \text{Re}(g_{eff}^{\uparrow\downarrow}) + \text{Im}(g_{eff}^{\uparrow\downarrow})$. In non-magnetic metal layer, the imaginary component of the complex effective spin-mixing conductance is significantly smaller than the real part[39, 45]. The real part, which primarily influences the efficiency of spin transfer across the FM/NM interfaces is closely associated with the intrinsic spin-mixing conductance in the following manner[43, 44]:

$$g_{eff}^{\uparrow\downarrow} = g_{int}^{\uparrow\downarrow}\left(1-e^{\frac{-t}{\lambda_{sd}}}\right) \quad (4)$$

As the effective spin mixing conductance offers valuable insights into the interfacial spin transfer characteristics it is an important parameter to consider when developing spintronic devices with one or more FM/NM layers.

In equation (3), the term $\alpha_{sp}$ is derived from $\alpha_{eff}$ which encapsulates all the contributions to the experimentally measured damping such as the intrinsic damping ($\alpha_{int}$) of the FM layer and the additional damping resulting from spin-pumping. $\alpha_{eff}$ can be given as follows[46]:

$$\alpha_{eff} = \alpha_{int} + \alpha_{sp} \quad (5)$$

We note that in addition to the spin-pumping in Pt the seed and capping NM layers also contribute to the spin-pumping. Therefore, to extract the spin-pumping damping in Pt we must separate these additional contributions. Accordingly, equation (5) can be re-written for the $Ni_{80}Fe_{20}$ and Co layers as the following equations (6) and (7) respectively.

$$\alpha_{eff,NiFe} = \alpha_{NiFe} + \alpha_{sp,seed} + \alpha_{sp,Pt} \qquad (6)$$
$$\alpha_{eff,Co} = \alpha_{Co} + \alpha_{sp,cap} + \alpha_{sp,Pt} \qquad (7)$$

Where, $\alpha_{NiFe}$ and $\alpha_{Co}$ represent the intrinsic damping of $Ni_{80}Fe_{20}$ and Co layers respectively, $\alpha_{sp,seed}$ and $\alpha_{sp,cap}$ represent the damping in seed and capping layers respectively and $\alpha_{sp,Pt}$ is the damping due to spin-pumping in Pt.

The general form can be written as follows:

$$\alpha_{eff} = \alpha_{ref} + \alpha_{sp,Pt} \qquad (8)$$

$\alpha_{ref}$ is the reference damping due to both the intrinsic and extrinsic sources. This equation thus accounts for the diverse sources of damping. We separately found $\alpha_{ref}$ for Ta(2)/$Ni_{80}Fe_{20}$(6)/Cu(3) and Co(6)/Ta(2) layers to be $8.21 \times 10^{-3}$ and $5.7 \times 10^{-3}$ respectively.

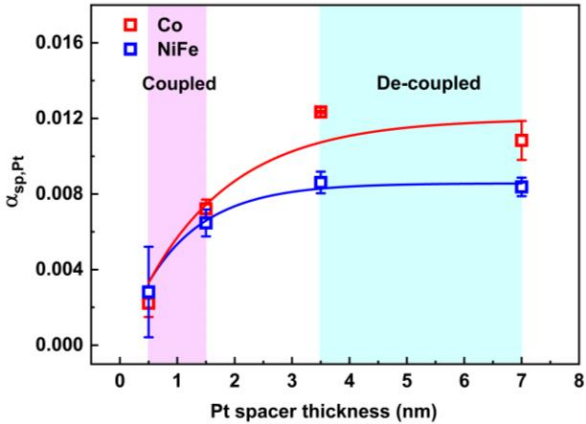

Figure 4: Spin-pumping induced damping at $Ni_{80}Fe_{20}$/Pt and Co/Pt interfaces as a function of spacer thickness, solid line represents the fit by spin-pumping model using equation (3) to extract the spin-diffusion length ($\lambda_{sd}$) and intrinsic spin-mixing conductance ($g^{\uparrow\downarrow}_{int}$).

The variation of $\alpha_{sp,Pt}$ as a function of Pt spacer thickness for $Ni_{80}Fe_{20}$ and Co is shown in Figure 4. The value of $\alpha_{sp,Pt}$ increases with $t_{Pt}$ and reaches different values of saturation for the two FM layers; $7.8 \times 10^{-3}$ for $Ni_{80}Fe_{20}$ and up to $12 \times 10^{-3}$ for Co. A higher value of damping for Co indicates larger transfer of spin angular momentum out of the precessing FM through its interface with adjacent NM[39] and a faster relaxation of precession in Co as compared to $Ni_{80}Fe_{20}$ in the decoupled regime. A mutual transfer of spin angular momentum between the two FM layers results in nearly equal relaxation rates in the coupled regime[17]. With a Cu(7) spacer, $\alpha_{sp,Pt}$, measured $4.3 \times 10^{-3}$ and $4.7 \times 10^{-3}$ for Co and $Ni_{80}Fe_{20}$ respectively, are lower than the values obtained for Pt. This difference can be attributed to the magnetic proximity effect at the FM/Pt interface[47]. The magnetic proximity effect refers to the induction of magnetic order by FM layer in neighbouring NM layer at their interface. It's effect at FM/Pt interface is significantly higher than that at FM/Cu interface because firstly Pt has stronger SOC than Cu and secondly the Pt has partially filled d-orbitals, compared to fully filled d-orbital for Cu, which allows greater interaction with d-orbital electrons of the FM layer. This result suggests that by considering the magnetic proximity effects at the FM/NM interface we can design spacer layers to tune the transfer of spin current in devices.

Solid lines in Figure 4 show the curve fitting of $\alpha_{sp,Pt}$ using equation (3) for $Ni_{80}Fe_{20}$ and Co. The derived values of $g^{\uparrow\downarrow}_{int}$ at the $Ni_{80}Fe_{20}$- and Co/Pt interfaces, are $1.73 \times 10^{19}(\pm 4.5 \times 10^{17})$ and $4.12 \times 10^{19}(\pm 5.34 \times 10^{18})$ respectively. While $\lambda_{sd}$ values of Pt are found to be $1.02 \times 10^{-9}(\pm 1.44 \times 10^{-10})$ and $1.55 \times 10^{-9}(\pm 6.05 \times 10^{-10})$ from the fitting of the $Ni_{80}Fe_{20}$ and Co data, respectively. These values lie in the range reported for bilayer and trilayer heterostructures of Pt as seen in Table I.

Using equation (4), we determined $g^{\uparrow\downarrow}_{eff}$, the values that we obtain are tabulated in Table and are found to be consistent with those reported in the literature. The variation of $g^{\uparrow\downarrow}_{eff}$ as a function of Pt thickness is shown in Fig. S.5 in Supplementary Information Section S-IV.

Table I: Estimated effective spin-mixing conductance and spin diffusion length at $Ni_{80}Fe_{20}$/Pt and Co/Pt, in the context of literature.

| $g^{\uparrow\downarrow}_{eff}$ $(m^{-2})$ ($Ni_{80}Fe_{20}$/Pt interface) | $g^{\uparrow\downarrow}_{eff}$ $(m^{-2})$ (Co/Pt interface) | $\lambda_{sd}$ $(m)$ |
|---|---|---|
| $1.72 \times 10^{19}$ **This work** | $4.07 \times 10^{19}$ **This work** | $1.02 \times 10^{-9} \pm 1.44 \times 10^{-10}$ ($Ni_{80}Fe_{20}$/Pt) $1.55 \times 10^{-9} \pm 6.05 \times 10^{-10}$ (Co/Pt) **This work** |
| $1.08 \times 10^{20}$ [48] | $3.96 \times 10^{19} \pm 3.9 \times 10^{18}$ [49] | $1.1 \times 10^{-9}$ [11] |
| $1.52 \times 10^{19} \pm 3.4 \times 10^{18}$ [49] | $4.5 \times 10^{19}$ [50] | $1.4 \times 10^{-9}$ [14, 51] |
| $2.1 \times 10^{19}$ [52] |  | $1.5 \times 10^{-9}$ [53] |
| $2.4 \times 10^{19}$ [54] |  | $2.4 \times 10^{-9}$ [55] |
| $4.0 \times 10^{19}$ [50] |  |  |
| $4.3 \times 10^{19} \pm 4 \times 10^{18}$ [11] |  |  |

Consistent with our findings, the phenomenon of non-reciprocal spin-pumping damping characterized by varying spin-mixing conductance values and their dependency on spacer thickness, has been identified in FeCoV/Ru/NiFe asymmetric trilayers as well, with higher spin-mixing conductance values reported for NiFe/Ru interface in trilayers ranging from $11.7 - 7.7) \times 10^{19} m^{-2}$ compared to NiFe/Ru bilayer ~ $(3.9 \pm 0.3) \times 10^{19} m^{-2}$ [15]. This variance underscores the significant role of trilayer configurations in modulating spin-mixing conductance. While recent literature has touched upon reciprocal spin-pumping damping in asymmetric trilayers[16], however, our observations distinctly demonstrate the non-reciprocal nature of $\alpha_{sp}$ across different FM layers, further emphasizing the complexity of spin interactions in such engineered heterostructures.

From the point of view of generation, manipulation and detection of spin currents for different spintronic applications, it is important to quantify the interfacial spin transparency ($T_{FM/Pt}$) for FM/Pt interface. For example, in magnetic memories, spin filter and spin valves understanding $T_{FM/Pt}$ can significantly enhance the efficiency, maximize the spin polarized current, etc. $T_{FM/Pt}$ essentially measures the impedance to spin current flow at the interface, influenced by electronic state mismatch and lattice imperfections. It is directly related to the effective spin mixing conductance ($g_{eff}^{\uparrow\downarrow}$) as follows[44, 49]:

$$T_{FM/Pt} = \frac{g_{eff}^{\uparrow\downarrow} tanh\left(\frac{t}{2\lambda}\right)}{g_{eff}^{\uparrow\downarrow} coth\left(\frac{t}{\lambda}\right) + \frac{h}{2\lambda e^2 \rho}} \quad (9)$$

where, $\rho$ (= 20 $\mu\Omega.cm$) [49], $t$ and $\lambda$ are the resistivity, thickness and spin-diffusion length of Pt. $T_{Co/Pt}$ is found to be ~ 62% and $T_{NiFe/Pt}$ ~ 25%. Both the values are in agreement with the reported values for Co/Pt and Ni$_{80}$Fe$_{20}$/Pt interfaces[49]. The higher $T_{FM/Pt}$ observed at the Co/Pt interface suggests a more conducive pathway for spin current flow in comparison to the Ni$_{80}$Fe$_{20}$/Pt interface.

Based on phenomenological model of spin-pumping, the interfacial spin current density ($j_s^{FM/Pt}$) has been related to the effective Gilbert damping parameter, $\alpha_{eff}$, and NM material dependent spin-mixing conductance $g_{eff}^{\uparrow\downarrow}(Pt)$[43, 56]:

$$j_s^{FM/Pt} = \frac{g_{eff}^{\uparrow\downarrow}(Pt)\gamma^2 h_{rf}^2 \hbar\left[+\sqrt{(4\pi M_s)^2\gamma^2 + 4\omega^2}\right]}{8\pi\alpha_{eff}^2[(4\pi M_s)^2\gamma^2 + 4\omega^2]} \quad (10)$$

here,

$$g_{eff}^{\uparrow\downarrow}(Pt) = g_{eff}^{\uparrow\downarrow}\left(1 + \left(2\sqrt{\varepsilon/3} tanh(t_{Pt}/\lambda)\right)^{-1}\right)^{-1} \quad (11)$$

where, $g_{eff}^{\uparrow\downarrow}$ is the effective spin-mixing conductance calculated using (4) and $\varepsilon = (Z_{Pt}e^2/\hbar c)^4 = 0.1$. Note that, $g_{eff}^{\uparrow\downarrow}$ does not account for the material properties of NM, in contrast to the $g_{eff}^{\uparrow\downarrow}(Pt)$, calculated using equation (11). The estimated $j_s^{FM/Pt}$ at the Co/Pt and Ni$_{80}$Fe$_{20}$/Pt interfaces is shown in Figure 5.

We see that the spin current density $j_s^{FM/Pt}$ increases with decreasing Pt thickness for both the interfaces, following the inverse dependence on $\alpha_{eff}$. This trend is indicative of the enhanced backflow of spins that occurs when Ni$_{80}$Fe$_{20}$ and Co layers are strongly coupled. These results can be related to the spin-pumping model discussed earlier, in that the backflow of spins is enhanced at the FM/NM interface because of the spin current pumped out of the opposing FM layer. This explains the observed increase in spin-current density at the FM/Pt interface with decreasing Pt thickness. Further, the higher value of spin current density at the Co/Pt interface can be attributed to its greater interfacial spin transparency of ~62% as compared to that of ~25% for Ni$_{80}$Fe$_{20}$/Pt interface, estimated earlier. The ability to modulate the spin current density by adjusting spacer thickness and material properties thus offers a dynamic control mechanism for spin interactions for an asymmetrical trilayer system.

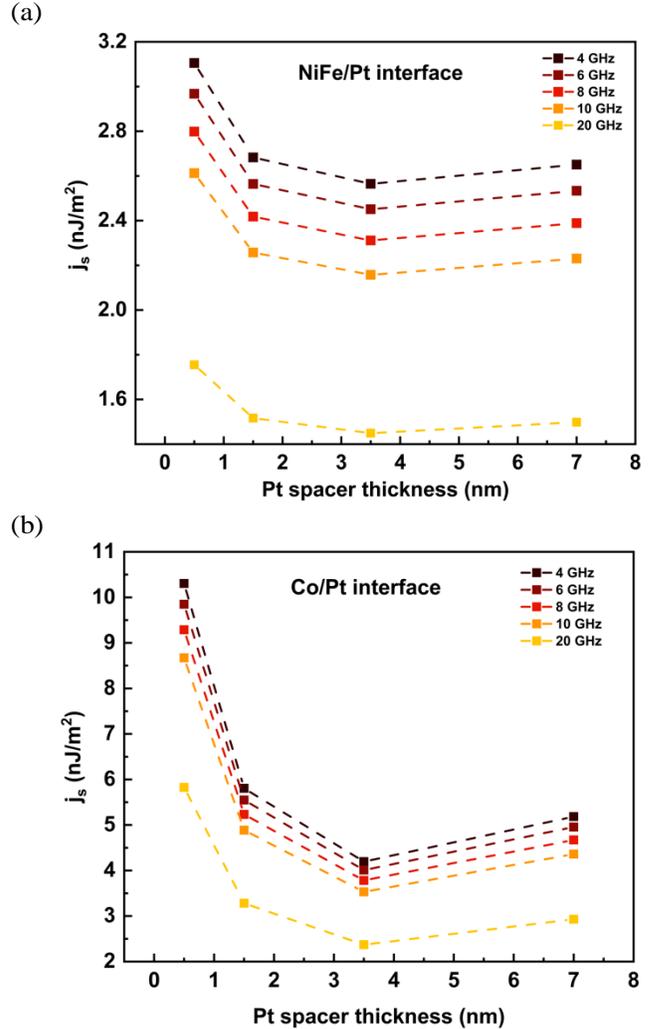

Figure 5: Spin current density at (a) Ni$_{80}$Fe$_{20}$/Pt and (b) Co/Pt interface as a function of Pt spacer thickness, for different rf excitation frequencies, calculated using (11).

## VI. PHASE OF PRECESSING MAGNETIZATIONS

We further elucidate our understanding of the spin-pumping mechanism in the trilayer by analyzing it as a composite of coupled FM layers that interact through both dynamic coupling via spin-pumping and static coupling via Pt-mediated exchange interactions.

To facilitate this analysis, we compute the average spin-pumping induced damping ($\tilde{\alpha}_{sp}$) within the trilayer system, using the experimentally obtained saturation magnetizations ($M_s$) of the two FM layers as the weighting factors, as considered in equation (12),

$$\tilde{\alpha}_{sp} = \frac{M_{s,Co} \times \alpha_{Co} + M_{s,NiFe} \times \alpha_{NiFe}}{M_{s,Co} + M_{s,NiFe}} \quad (12)$$

with, $M_{s,Co} = 1459$ emu/cc and $M_{s,NiFe} = 825$ emu/cc, alongside the experimentally determined Gilbert damping parameters, $\alpha_{Co}$ and $\alpha_{NiFe}$. The variation of $\tilde{\alpha}_{sp}$ as a function of $t_{Pt}$ is shown in Figure 6.

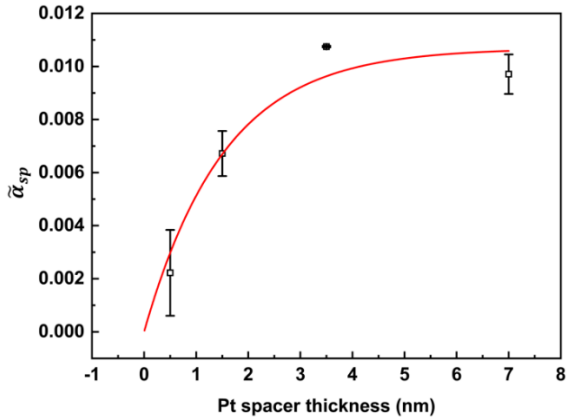

Figure 6: Average spin-pumping induced damping of the acoustic mode as a function of Pt spacer thickness. Solid red line represents fit to the data using equation (13).

The data is found to fit reasonably well to the following equation for spin-pumping induced damping of the acoustic mode[11]:

$$\tilde{\alpha}_{sp} = \frac{g \mu_B}{4\pi M_s} \frac{\tilde{g}_{\uparrow\downarrow}}{t_{FM}} \left[ 1 + \frac{\tilde{g}_{\uparrow\downarrow} \mathcal{R}}{\tanh\left(\frac{t_{Pt}}{2\lambda_{sd}}\right)} \right]^{-1} \quad (13)$$

where, $\mathcal{R} = \frac{\rho_\uparrow e^2}{2\pi\hbar} \lambda_{sd}$, $\rho_\uparrow (= 34 \pm 1 \ \mu\Omega.cm)$ [11] is the single spin channel resistivity, $t_{FM}$ is the thickness of the FM and $g = \frac{M_{s,Co} \times g_{Co} + M_{s,NiFe} \times g_{NiFe}}{M_{s,Co} + M_{s,NiFe}} = 2.15$, where $g_{Co}$ and $g_{NiFe}$ represent the g-factors of Co and Ni$_{80}$Fe$_{20}$, respectively. The 'composite' spin-mixing conductance and spin-diffusion length given by $\tilde{g}_{\uparrow\downarrow}$ and $\lambda_{sd}$ respectively are the fitting parameters. The data shows a good fit to the model, suggesting that the coupled FM layers undergo in-phase acoustic mode precession. Unlike similar analyses for symmetric NiFe/Pt/NiFe trilayers[11], our asymmetric interfaces exhibit distinct behavior with a 'composite' spin-mixing conductance, $\tilde{g}_{\uparrow\downarrow} = 5.43 \times 10^{19} \pm 1.42 \times 10^{18} \ m^{-2}$ and spin-diffusion length, $\tilde{\lambda}_{sd} = 1.49 \times 10^{-9} \pm 1.91 \times 10^{-10} \ m$, indicating the complex magnetization dynamics at play in an asymmetrical trilayer system.

As outlined in the introduction, magnetization precession in the coupled FM layers emits spin current out of the FM layers with parallel or opposite spin polarization, contingent upon the phase of the precessing magnetization. For NM spacer thicknesses below $\lambda_{sd}$, the emitted spin current traverses the conducting Pt spacer layer via ballistic transport and is absorbed by the other FM layer[10]. Notably, for in-phase precession and $t_{Pt} < \lambda_{sd}$, the loss of angular momentum is mitigated by the absorption of incoming spin angular momentum, resulting in reduced spin-pumping induced damping as the spacer thickness diminishes[11]. This mechanism facilitates spin-pumping induced dynamic coupling between the two layers, alongside the Pt-mediated static exchange coupling. The combination of dynamic and static coupling between the FMs contributes to the observed enhancement in linewidth and shifts in the resonance field of the coupled FM layers as a function of Pt spacer thickness.

## VII. CONCLUSION

In summary, our systematic study of Ni$_{80}$Fe$_{20}$/Pt(t)/Co trilayer system demonstrates that variation in the Pt NM spacer thickness significantly influences the coupling of FM layers, as analyzed through VSM and macrospin simulations. An exponential decrease in the coupling strength with increasing Pt thickness is observed. Subsequently, the trilayer system with strongly coupled FM layers, with 0.5 nm and 1.5 nm Pt spacer layer, exhibits a single resonance peak whereas in the systems with decoupled FM layers, for thicker Pt spacer layer (3 nm and 7 nm), two well separated resonance peaks emerge, corresponding to decoupled resonances in Ni$_{80}$Fe$_{20}$ and Co respectively. Spin-pumping in Ni$_{80}$Fe$_{20}$/Pt(t)/Co studied by FMR reveals a nuanced dependency of spin-pumping induced damping ($\alpha_{sp}$) on the Pt spacer thickness, highlighting distinctive non-reciprocal behaviours at the Ni$_{80}$Fe$_{20}$/Pt and Co/Pt interfaces. The observed variation in the saturation levels of $\alpha_{sp}$ for Ni$_{80}$Fe$_{20}$ and Co is indicative of the differential spin transport characteristics inherent to each FM/Pt interface. The effective $\alpha_{sp}$ is found to decrease with decreasing Pt thickness, indicating a dynamic compensation of the spin current achieved through the in-phase precession of magnetizations in Ni$_{80}$Fe$_{20}$ and Co, because of the in-phase spin current pumped by the other FM layer across the thin NM spacer layer. This observation, coupled with the different values of spin-mixing conductances at the two interfaces, underscores the substantial difference in the spin-current density present at two interfaces in the asymmetric trilayer system. The derived interfacial spin transparency, approximately ~62% for Co/Pt and ~25% for Ni$_{80}$Fe$_{20}$/Pt interfaces, alongside the measured spin current densities of 10.5 $nJ/m^{-2}$ and 3.2 $nJ/m^{-2}$ respectively, points out the differential spin transport efficacy across these interfaces.

This investigation establishes the foundation for addressing the theoretical integration of coupling strength and dynamic spin-pumping between the FM layers for the analysis of $\alpha_{sp}$ in an asymmetric trilayer system. By developing a fundamental understanding of spin-pumping in an asymmetric trilayer and elucidating the function of Pt in regulating the spin transfer dynamics this study provides a framework for future research into the mechanism of spin transport and targeted advancements in spintronic device design to improve the efficiency and performance of future applications.

## ACKNOWLEDGEMENTS

S.S. is thankful to Dr. Durgesh Kumar for discussion related to VSM experimental results. S.S. acknowledges receiving PhD Scholarship through NRF-CRP21-2018-0003 grant from National Research Foundation (NRF) Singapore. R.S.R. acknowledges the support from NRF, Singapore through NRF-CRP21-2018-0003 grant and the Ministry of Education, Singapore through its Academic Research Tier 1 grant, RG76/22 and Tier 2 grant MOE-T2EP50122-0023. Any opinions, findings and conclusions or recommendations expressed in this material are those of the author(s) and do not reflect the views of the Ministry of Education, Singapore. Y.R. and Y.M. are supported by the National Natural Science Foundation of China (Grant No. 62074099).

# Supplementary Information

### S-I: Features of the M-H hysteresis loops

For NM spacer layer thicknesses t ≥ 3 nm, the M-H hysteresis loops display a two-step transition. This behaviour indicates that the magnetization vectors of the two FM layers switch independently, signifying a weak coupling or effective decoupling between the layers.

The influence of coupling between the ferromagnetic (FM) layers significantly manifests in the intrinsic coercive field ($H_{ic}$) of the samples. Notably, the $H_{ic}$ values for the individual $Ni_{80}Fe_{20}$ and Co layers are determined to be 2 Oe and 20 Oe, respectively. However, for samples exhibiting strong coupling between FM layers, the observed $H_{ic}$ ranges between 10 Oe to 14 Oe. This intermediate coercivity underscores the combined influence of the magnetic properties of the individual FM layers and the nature of their interlayer coupling within the $Ni_{80}Fe_{20}$/Pt/Co trilayers, particularly when employing thin non-magnetic (NM) spacers. Notably, for spacer thicknesses t ≥ 3 nm, the hysteresis loops display a two-step transition. This behaviour indicates that the magnetization vectors of the two FM layers switch independently, signifying a weak coupling or effective decoupling between the layers. Additionally, the switching field for the Co layer in the Pt(7) sample is observed to be approximately 34 Oe, compared to roughly 19 Oe in the sample with a Cu(7) spacer, highlighting the significant role of spacer material and thickness in modulating the magnetic interactions and properties of $FM_1$/NM/$FM_2$ trilayers. Furthermore, this enhancement in coercivity can be attributed to the formation of PMA induced by the Co-Pt interfaces leading to higher anisotropy.

Additionally, the M-H hysteresis loops display a gradual slope rather than a sharp switch in magnetization, simulations provide insight into this behavior suggesting that the observed slope during magnetization reversal can be ascribed to the trilayer's non-ideal easy axis alignment. This phenomenon, along with the previously noted premature magnetization reversal, supports the presence of dipole-dipole coupling. Such coupling impacts the alignment of magnetic moments between adjacent ferromagnetic (FM) layers, affecting the trilayer's magnetic properties. The variations in saturation magnetization ($M_s$) as a function of Pt spacer thickness is shown in Fig. S.1. For spacer thickness greater than 1.5 nm $M_s$ is found to saturate around 1150 emu/cc. $M_s$ is observed to be enhanced after inclusion of even thin Pt(0.5) spacer as compared to Pt(0), further it reaches a saturation for continuous Pt films of higher thickness. The enhancement can be attributed to the induction of magnetic order in Pt by the neighbouring FM layers while the saturation indicates the maximum induced magnetic order in Pt.

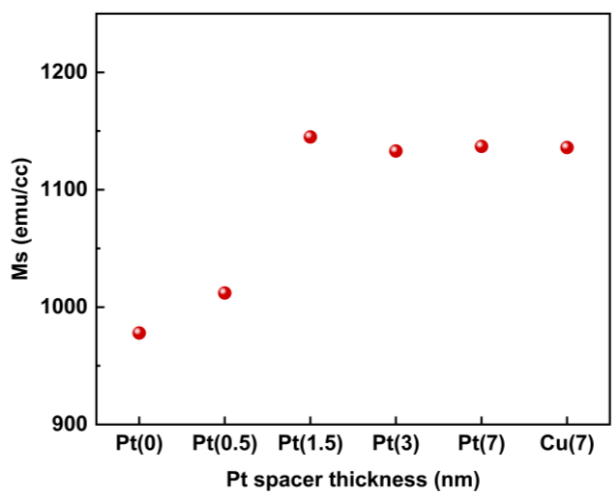

Fig. S.1: Variation in saturation magnetization as a function of Pt spacer thickness.

## S-II: Details and results of macrospin model

To better understand the exchange coupling of the trilayer system, macro-spin model is used to describe the magnetization rotation of the uniform individual ferromagnet layer. For thin films with in-plane anisotropy, the free energy density per unit area can be expressed as[1]:

$$E_n(m_i) = E_{Zeeman} + E_{Dem} + E_k + E_{ex} \quad (S.1)$$

$$= -\mu_0 H M_s m_1 + \frac{\mu_0}{2}(M_s m_1 \cdot \vec{z})^2 - K_u(m_1 \cdot \vec{x})^2 - \mu_0 H M_s m_2 + \frac{\mu_0}{2}(M_s m_2 \cdot \vec{z})^2 - K_u(m_2 \cdot \vec{x})^2$$

$$- \frac{J_1}{d} m_1 \cdot m_2 + \frac{J_2}{d}(m_1 \cdot m_2)^2$$

where, $E_{Zeeman}$, $E_{Dem}$, $E_k$, $E_{ex}$ are the Zeeman, demagnetization, anisotropy and exchange energy terms respectively and $\mu_0$, $M_s$, $K_u$, $J_1$, $J_2$ and $d$ are the vacuum permeability, saturation magnetization, in-plane uniaxial anisotropy constant, bilinear coupling, biquadratic coupling and ferromagnet thickness, respectively. Note that when $J_1 \gg J_2$ and $J_1 > 0$ corresponds to the ferromagnetic coupling of adjacent layers. $m_i$ denotes the unit magnetization vector for the *i*-th ferromagnet layer and can be written as $m_1 = (\cos\varphi_1, \sin\varphi_1, 0)$ and $m_2 = (\cos\varphi_2, \sin\varphi_2, 0)$. However, the tilted experimental hysteresis loops at the switching fields instead of sharp transition indicate that the easy-axis of the sample is not ideal. The results with non-ideal easy axis can be simulated closer to the experiment by using a Gaussian function $\frac{1}{\sigma\sqrt{2\pi}} exp\left(\frac{(H-H_0)^2}{2\sigma^2}\right)$ that controls the easy axis distribution, where $1/\sigma$, $H$ and $H_0$ are the standard deviation of the Gaussian function, magnetic field in any direction and magnetic field along easy axis. A higher value of sigma indicates a lower deviation from ideal easy axis. Table S.I details the values of $1/\sigma$, $J_1$ and $J_2$.

Table S. I: the standard deviation, bilinear and biquadratic coupling strength for different spacers as extracted from simulations of the M-H hysteresis loops.

| Spacer(t) nm | $1/\sigma$ | $J_1\ (erg/cm^{-2})$ | $J_2\ (erg/cm^{-2})$ |
|---|---|---|---|
| Pt(0) | 0.25 | 10 | 0.1 |
| Pt(0.5) | 1 | 4 | 0.1 |
| Pt(1.5) | 0.6 | 0.028 | 0.1 |
| Pt(3) | 1 | 1x10$^{-4}$ | < 1x10$^{-3}$ |
| Pt(7) | 0.6 | 0 | 0 |
| Cu(7) | 0.6 | 0 | 0 |

Fig. S.2 shows the simulated and experimental VSM data for samples with varying Pt spacer thickness and the Cu spacer.

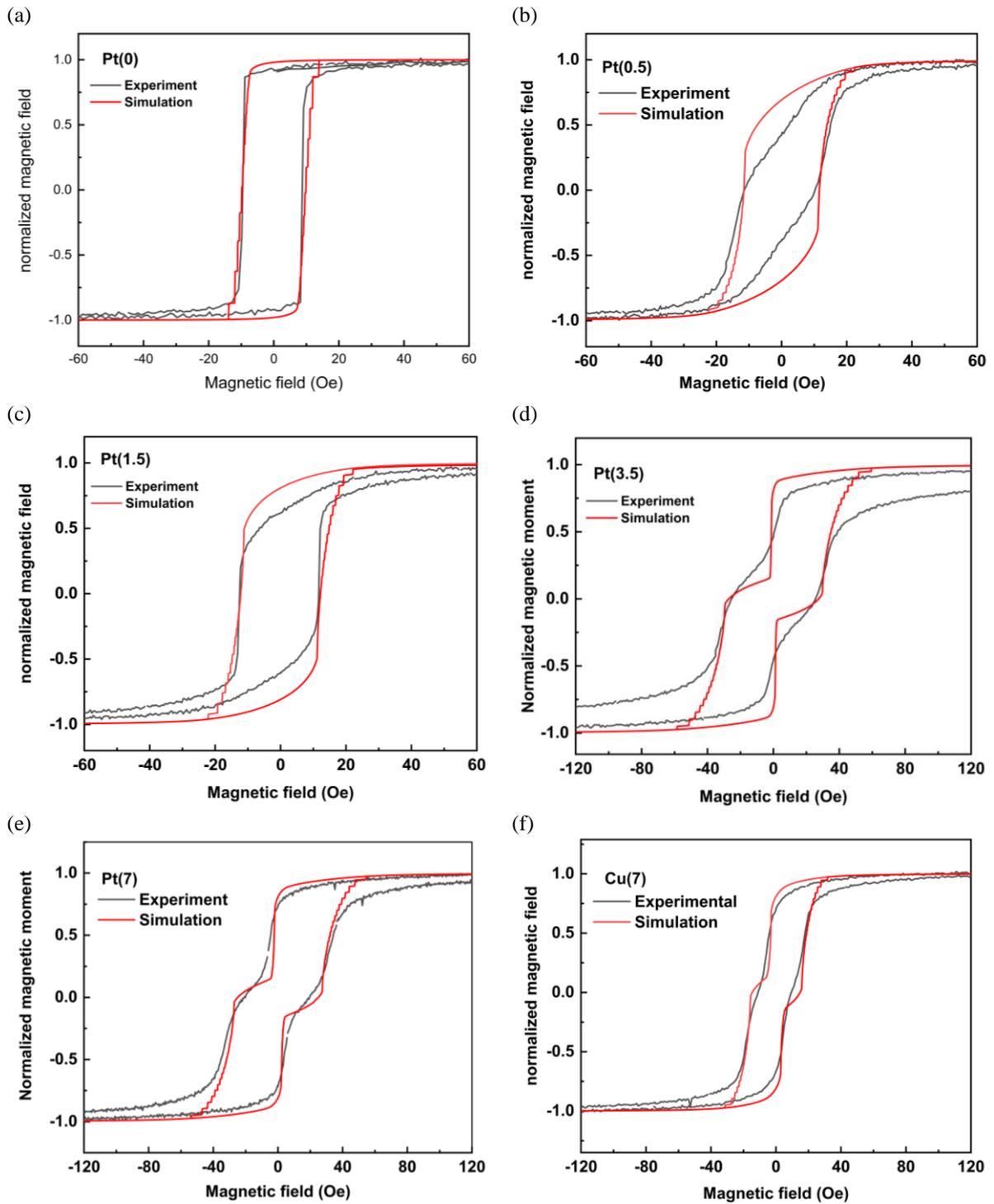

Fig. S.2: Simulated (in red) and experimental (in black) M-H hysteresis data for varying Pt spacer thickness (a)-(e) and Cu(7) spacer (f).

## S-III: Analysis of FMR spectra

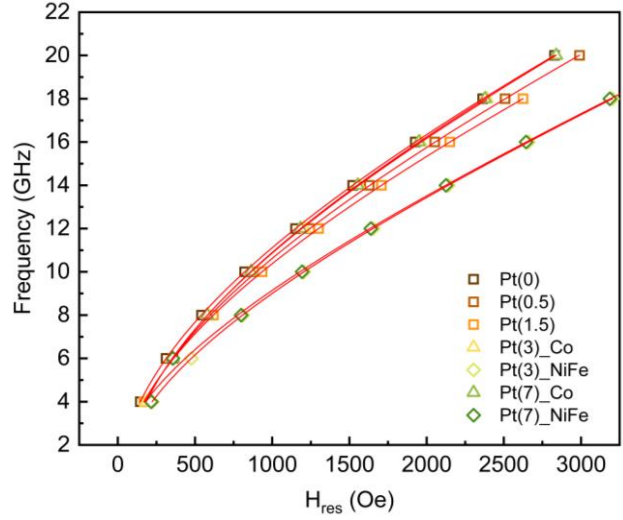

Fig. S.3: RF excitation frequency vs. resonance field (H$_{res}$), solid lines are fits using the Kittel equation using equation (2). For decoupled trilayers data for both NiFe and Co layers is shown.

The effective magnetization ($4\pi M_{eff}$) was calculated by fitting the frequency vs. $H_{res}$ data, shown in Figure S.3, to the Kittel's equation for in-plane configuration of FMR as described by equation (S.2). This equation is obtained by solving the LLG equation in the small-angle precession limit of magnetization.

$$f = \frac{\gamma}{2\pi}\left[(H_{res} + H_k)(H_{res} + H_k + 4\pi M_{eff})\right]^{1/2} \tag{S.2}$$

Where, f represents the resonance frequency, $\gamma$ the gyromagnetic ratio with $\gamma = g\mu_B/\hbar = g \times 87.94$ Hz/T, $H_{res}$ the resonance field and $H_k$ is the in-plane anisotropy field. Given typical values for Co and NiFe g-factors as 2.18 and 2.1 respectively, we observed that in weakly coupled or decoupled samples ($t_{Pt} = 3$ and 7 nm and $t_{Cu} = 7$ nm) the effective magnetization ($4\pi M_{eff}$) for the initial peak at lower magnetic field and the subsequent peak at higher magnetic field measure approximately 12-13.5 kOe and 8.1 kOe respectively as shown in Fig. 3 of the main text. The effective magnetization in the coupled regime is the weighted average with weights as the saturation magnetization of Co and NiFe. These findings distinctly attribute the initial and subsequent peaks to the ferromagnetic resonances of Co and NiFe, respectively. From Fig. S.4 it is evident that in the coupled FM layers, the effective magnetization decreases with an increase in Pt spacer thickness. In the decoupled regime, as expected, effective magnetization of the Co and NiFe layers revert to the aforementioned values.

For thin films, the interfacial anisotropy is larger than the bulk anisotropy and it incorporates both shape anisotropy and in-plane crystalline anisotropy, we determine the value of $H_\perp$ from the following relation, $H_\perp = 4\pi M_s - 4\pi M_{eff}$[2, 3], where the values of saturation magnetization ($M_s$) and $4\pi M_{eff}$ are extracted from VSM and FMR measurements respectively. The enhancement in the interfacial uniaxial magnetic anisotropy field perpendicular to the film plane ($H_\perp$) with increasing Pt thickness in the strongly coupled regime as shown in the inset of Fig. S.4 is linked to the observed decrease in $4\pi M_{eff}$. This enhancement at the Co/Pt interface stems from the Co 3d-Pt 5d interfacial hybridization, a well-documented effect contributing to the increased perpendicular magnetic anisotropy[4, 5]. When the Pt spacer thickness exceeds 1.5 nm, the perpendicular anisotropy field doesn't change with further increase in Pt thickness and the effective magnetization ($4\pi M_{eff}$) of Co remains nearly constant. Conversely, the control sample with 7 nm thick Cu spacer, demonstrates a significant reduction in $H_\perp$, attributable to the disfavor of Cu/Co interface toward out-of-plane magnetic moment alignment. This substantial decrease further supports our explanation of the observed reduction in $4\pi M_{eff}$ of strongly coupled samples.

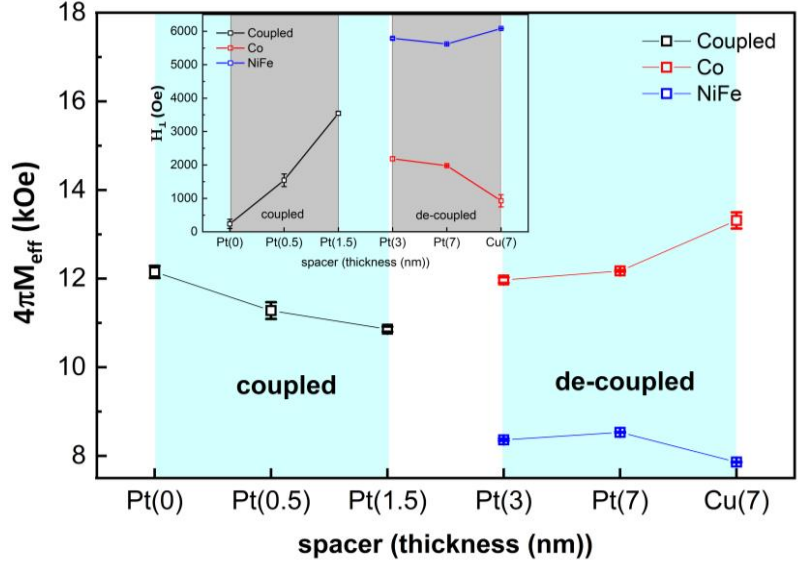

Fig. S.4: Effective magnetization ($4\pi M_{eff}$) as a function of Pt spacer thickness, inset shows the calculated interfacial anisotropy ($H_\perp$). Also, shown is the data for the Cu(7) control sample for comparison. Solid line is a guide to the eye.

### S-IV: Effective spin-mixing conductance as a function of Pt spacer thickness

Fig. S.5 shows the variation in the effective spin-mixing conductance ($g_{eff}^{\uparrow\downarrow}$) of the trilayers, calculated using equation (4), as a function of Pt spacer thickness for $Ni_{80}Fe_{20}$/Pt and Co/Pt interfaces. It indicates change in the efficiency of spin current with respect to the spacer thickness. As expected from equation (4), $g_{eff}^{\uparrow\downarrow}$ increases with Pt thickness and for $t_{NM} \gg \lambda_{sd}$, its value saturates at $g_{int}^{\uparrow\downarrow}$. This result shows how the spacer material approaches ideal interface characteristics as the spin-backflow is compensated by the spin scattering in Pt.

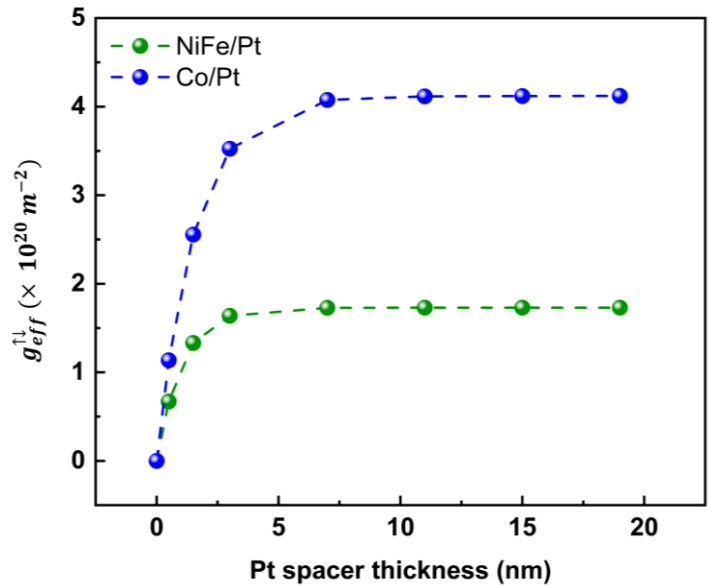

Fig. S.5: effective spin-mixing conductance as a function of Pt spacer thickness calculated using equation (4). Dashed line is guide to the eye.